# Two-dimensional Nanolithography Using Atom Interferometry

A. Gangat, P. Pradhan, G. Pati, and M.S. Shahriar

*Department of Electrical and Computer Engineering*

*Northwestern University,*

*Evanston, IL 60208*

**Abstract**

We propose a novel scheme for the lithography of *arbitrary*, *two-dimensional* nanostructures via matter-wave interference. The required quantum control is provided by a $\pi/2$- $\pi$ - $\pi/2$ atom interferometer with an integrated atom lens system. The lens system is developed such that it allows simultaneous control over atomic wave-packet spatial extent, trajectory, and phase signature. We demonstrate arbitrary pattern formations with two-dimensional $^{87}$Rb wave-packets through numerical simulations of the scheme in a practical parameter space. Prospects for experimental realizations of the lithography scheme are also discussed.

# I. INTRODUCTION

The last few decades have seen a great deal of increased activity toward the development of a broad array of lithographic techniques [1, 2]. This is because of their fundamental relevance across all technological platforms. These techniques can be divided into two categories: parallel techniques using light and serial techniques using matter. The optical lithography techniques have the advantage of being fast because they can expose the entire pattern in parallel. However, these techniques are beginning to reach the limits imposed upon them by the laws of optics, namely the diffraction limit [3]. The current state-of-the-art in optical lithography that is used in industry can achieve feature sizes on the order of hundreds of nanometers. Efforts are being made to push these limits back by using shorter wavelength light such as x-rays [2], but this presents problems of its own. The serial lithography techniques, such as electron beam lithography [1], can readily attain a resolution on the order of tens of nanometers. However, because of their serial nature these methods are very slow and do not provide a feasible platform for the industrial mass fabrication of nano-devices.

A new avenue for lithography presents itself out of recent developments in the fields of atomic physics and atom optics, namely the experimental realization of a Bose-Einstein Condensate (BEC) [4, 5] and the demonstration of the atom interferometer [6-12]. In essence, these developments provide us with the tools needed in order to harness the wave nature of matter. This is fortuitous for lithography because the comparatively smaller de Broglie wavelength of atoms readily allows for a lithographic resolution on the nanometer scale. The atom interferometer provides a means of interfering matter waves in order to achieve lithography on such a scale. The BEC, on the other hand, provides a highly coherent and populous source with which to perform this lithography in a parallel fashion. The opportunity thus presents itself to combine the enhanced resolution of matter interferometry with the high throughput of traditional optical lithography.

It should be noted that, although there has been research activity on atom lithography [13-15] for a number of years, most of the work has involved using standing waves of light as optical masks for the controlled deposition of atoms on a substrate. The primary limitations of using such optical masks are that the lithographic pattern can not be arbitrary and that the resolution of the pattern is limited to the 100nm scale. Since our scheme uses the atom interferometer, however, it allows for pattern formation by self-interference of a matter wave, and is thus unhampered by the inherent limitations of the optical mask technique.

In this paper we seek to demonstrate theoretically the use of the atom interferometer as a platform for nanolithography by proposing a technique that allows for the manipulation of a single atom wavepacket so as to achieve two-dimensional lithography of an arbitrary pattern on the single nanometer scale. To do this our scheme employs a lens system along one arm of the interferometer that performs Fourier imaging [3] of the wavepacket component that travels along that arm. By investigating such a technique for a single atom wavepacket, we hope to establish the viability of using a similar technique for a single BEC wavepacket, which would allow for truly high throughput lithography.

The paper is organized as follows. Section II presents an overview of the proposed technique. Sec. III and IV provide a theoretical analysis of the atom interferometer itself and our proposed imaging system, respectively. Sec. V is devoted to some practical considerations of the setup and its parameter space, and Sec. VI gives the results of numerical simulations. Finally, we touch upon the issue of replacing the single atom wavepacket with the macroscopic wavefunction of a BEC in Sec. VII. Appendices A and B show some of the steps in the derivations.

## II.A. Principles of Operation

In a π/2- π - π/2 atom interferometer (AI), which was first theoretically proposed by Borde [6] and experimentally demonstrated by Kasevich and Chu [7], an atom beam is released from a trap and propagates in free space until it encounters a π/2 pulse, which acts as a 50/50 beam splitter [16-22]. The split components then further propagate in freespace until they encounter a π pulse, which acts as a mirror so that the trajectories of the split beam components now intersect. The beams propagate in free space again until they encounter another π/2 pulse at their point of intersection, which now acts as a beam mixer. Because of this beam-mixing, any phase shift $\phi$ introduced between the beams before they are mixed will cause an interference to occur such that the observed intensity of one of the mixed beams at a substrate will be proportional to $1+\cos\phi$, much like the Mach-Zehnder interferometer [23] from classical optics. For our scheme we propose the same type of interferometer, but with a single atom released from the trap instead of a whole beam.

Now, if we introduce an arbitrary, spatially varying phase shift $\phi(x,y)$ between the two arms of the interferometer before they mix, the intensity of their interference pattern as observed on a substrate will be proportional to $1+\cos\phi(x,y)$. Thus, in our system, we use a fortuitous choice of $\phi(x,y)$ in order to form an arbitrary, two-dimensional pattern. This quantum phase engineering (already demonstrated for BECs [24, 25]) is achieved by using the ac-stark effect so that $I(x,y) \propto \phi(x,y)$, where $I(x,y)$ is the intensity of an incident light pulse.

Also, in order to achieve interference patterns on the nanoscale, $\phi(x,y)$ must itself be at nanometer resolution. However, reliable intensity modulation of a light pulse is limited to the sub-micron range due to diffraction effects. One way to address this is by focusing the wavepacket after it is exposed to the sub-micron resolution phase shift $\phi(x,y)$, thereby further scaling down $\phi(x,y)$ to nanometer resolution after it is applied to the wavepacket. Our scheme achieves this scaling via an atom lens system.

Additionally, just as with a gaussian laser beam, exposing a single gaussian wavepacket to a spatially varying phase shift $\phi(x,y)$ will cause it to scatter. In order for both the phase-shifted and non-phase-shifted components of the wavepacket to properly interfere, our lens system is also used to perform Fourier imaging such that, at the substrate, the phase-shifted component of the wavepacket is an unscattered gaussian that is properly aligned with its non-phase-shifted counterpart and has the phase information $\phi(x,y)$ still intact. Indeed, the lens system, which is created using the ac-stark effect, serves the double purpose of scaling down the phase information $\phi(x,y)$ from sub-micron resolution to single nanometer resolution and neutralizing the wavepacket scattering caused by the same phase shift $\phi(x,y)$.

## II. B. Schematic

In our overall scheme, represented by Fig. 1, the atoms are treated as lambda systems [26-33] (Inset B) and are prepared in ground state $|1\rangle$. A single atom trap [34-36] is used to release just one atomic wavepacket along the z-axis. After traveling a short distance, the wavepacket is split by a π/2 pulse into internal states $|1\rangle$ and $|3\rangle$. The state $|3\rangle$ component gains additional momentum along the y-axis and separates from the state $|1\rangle$ component after they both travel further along the z-axis. Next, a π pulse causes the two components to transition their internal states and thereby reflect their trajectories. The component along the top arm is now in the original ground state $|1\rangle$ and proceeds to be exposed to the lens system. The lenses of the lens system are pulses of light that intercept the state $|1\rangle$ component of the wavepacket at different times. By modulating their spatial intensity in the x-y plane, these pulses of light are tailored to impart a particular phase pattern in the x-y plane to the wavepacket component that they interact with via the ac-stark effect. As shown in Inset B, the detuning of the light that the lenses are composed of is several times larger for state $|3\rangle$ than for state $|1\rangle$. The lenses can therefore be considered to have a negligible ac-stark effect on the state $|3\rangle$ wavepacket component as compared to the state $|1\rangle$ component. The first light pulse is intensity modulated to carry the phase information of the first lens of the lens system. It then intercepts the state $|1\rangle$ wavepacket component and adds the phase $\phi_1(x,y)$. After some time the state $|1\rangle$ component evolves due to the first lens such that it is an appropriate size for exposure to the phase information corresponding to the arbitrary pattern image (Inset A). Another light pulse is intensity modulated to carry the phase information of both the second lens and the inverse cosine of the arbitrary pattern. The pulse intercepts the state $|1\rangle$ component and adds the additional phase $\phi_2(x,y)$. After some time a third light pulse is prepared and exposed to the state $|1\rangle$ component to add a phase of $\phi_3(x,y)$ to effect the third lens of the lens system. Soon after, the final $\pi/2$ pulse mixes the trajectories of the wavepacket components. After the light pulses for the lens system have passed through, a chemically treated wafer is set to intercept the state $|1\rangle$ component in the x-y plane. Due to the mixing caused by the last $\pi/2$ pulse, only part of what is now the state $|1\rangle$ component has gone through the lens system. Because of the lens system, it arrives at the wafer with a phase that is a scaled down version of the image phase $\phi_P(x,y) = \arccos P(x,y)$. The other part of what is now the state $|1\rangle$ component did not go through the lens system. There is therefore a phase difference of $\phi_P(x,y)$ between the two parts of the state $|1\rangle$ component and the wavepacket strikes the wafer in an interference pattern proportional to $1+\cos(\arccos P(x,y)) = 1+P(x,y)$. The impact with the wafer alters the chemically treated surface, and the pattern is developed through chemical etching.

As a note, one preparation for the wafer is to coat it with a self-assembled monolayer (SAM) [37]. However, S. B. Hill *et al.* [38] demonstrate an alternate approach using hydrogen passivation, which may be better suited for lithography at the single nanometer scale due to its inherent atomic-scale granularity.

### III. Analysis of the Interferometer ($\pi/2$ - $\pi$ - $\pi/2$)

**III. A. Formalism**

As explained in the previous section, we consider the behavior of a single atomic wavepacket in our formulation of the problem. Also, in order to understand and simulate the AI [6-12] properly, the atom must be modeled both internally and externally. It is the internal evolution of the atom while in a laser field that allows for the splitting and redirecting of the beam to occur in the AI. However, the internal evolution is also dependent on the external state. Also, while the external state of the atom accounts for most of the interference effects which result in the arbitrary pattern formation, the internal state is responsible for some nuances here as well.

In following the coordinate system as shown in Fig. 1, we write the initial external wavefunction as:

$$|\Psi_e(\vec{r}, t=0)\rangle = \frac{1}{\sigma\sqrt{\pi}} \exp\left(\frac{-|\vec{r}|^2}{2\sigma^2}\right) \tag{1}$$

where $\vec{r} = x\hat{i} + y\hat{j}$. Our use here of a two-dimensional model is justified because no measurement is made in the $z$ direction. Internally, the atom is modeled as a three level lambda system [26-33] (as shown in Fig. 1 Inset B) and is assumed to be initially in state $|1\rangle$:

$$|\Psi_i(t)\rangle = c_1(t)|1\rangle + c_2(t)|2\rangle + c_3(t)|3\rangle, \tag{2}$$

where we consider $c_1(0) = 1$, $c_2(0) = 0$, $c_3(0) = 0$. States $|1\rangle$ and $|3\rangle$ are metastable states, while state $|2\rangle$ is an excited state.

As will become evident later, in some cases it is more expedient to express the atom's wavefunction in k-space [39]. To express our wavefunction, then, in terms of momentum, we first use Fourier theory to re-express the external wavefunction as:

$$|\Psi_e(x, y, t)\rangle = \frac{1}{2\pi} \iint \Phi_e(p_x, p_y, t)|p_x\rangle|p_y\rangle dp_x\, dp_y, \tag{3}$$

where we let $|p_x\rangle = e^{i\frac{p_x}{\hbar}x}$ and $|p_y\rangle = e^{i\frac{p_y}{\hbar}y}$. The complete wavefunction is simply the outer product of the internal and external states (Eq. (2) and (3)):

$$|\Psi(x, y, t)\rangle = \frac{1}{2\pi} \iint [C_1(p_x, p_y, t)|1, p_x, p_y\rangle + C_2(p_x, p_y, t)|2, p_x, p_y\rangle \\ + C_3(p_x, p_y, t)|3, p_x, p_y\rangle] dp_x\, dp_y, \tag{4}$$

where $C_n(p_x, p_y, t) = c_n(t)|\Phi_e(p_x, p_y, t)\rangle$. In position space, the outer product gives:

$$|\Psi(\vec{r}, t)\rangle = c_1(t)|1, \Psi_e(\vec{r}, t)\rangle + c_2(t)|2, \Psi_e(\vec{r}, t)\rangle + c_3(t)|3, \Psi_e(\vec{r}, t)\rangle. \tag{5}$$

**III. B. State Evolution in Freespace**

The freespace evolution of a the wavefunction is fully derived in appendix A. Presented here are simply the results cast in our particular formalism. For the freespace Hamiltonian $H = \iint \sum_{n=1}^{3} \left( \frac{p_x^2 + p_y^2}{2m} + \hbar \omega_n \right) |n, p_x, p_y\rangle \langle n, p_x, p_y| dp_x dp_y$, if the wavefunction is known at time $t = 0$, then after a duration of time $T$ in freespace, the wavefunction becomes:

$$|\Psi(\vec{r}, t = T)\rangle = \frac{1}{2\pi} \iint \left[ C_1(p_x, p_y, 0) e^{-i\left(\frac{p_x^2 + p_y^2}{2m\hbar} + \omega_1\right)T} |1, p_x, p_y\rangle \right.$$
$$+ C_2(p_x, p_y, 0) e^{-i\left(\frac{p_x^2 + p_y^2}{2m\hbar} + \omega_2\right)T} |2, p_x, p_y\rangle$$
$$\left. + C_3(p_x, p_y, 0) e^{-i\left(\frac{p_x^2 + p_y^2}{2m\hbar} + \omega_3\right)T} |3, p_x, p_y\rangle \right] dp_x dp_y. \quad (6a)$$

We can also write it as:

$$|\Psi(\vec{r}, t = T)\rangle = e^{-i\omega_1 T} c_1(0) |1, \Psi_e(\vec{r}, T)\rangle + e^{-i\omega_2 T} c_2(0) |2, \Psi_e(\vec{r}, T)\rangle$$
$$+ e^{-i\omega_3 T} c_3(0) |3, \Psi_e(\vec{r}, T)\rangle \quad (6b)$$

**III. C. State Evolution in π and π/2 pulse laser fields**

The electromagnetic fields encountered by the atom at points 2, 3, and 7 in Fig. 1 that act as the π and π/2 pulses are each formed by two lasers that are counter-propagating in the y-z plane parallel to the y axis. We use the electric dipole approximation to write the hamiltonian in these fields as

$$H = \iint \sum_{n=1}^{3} \left( \frac{p_x^2 + p_y^2}{2m} + \hbar \omega_n \right) |n, p_x, p_y\rangle \langle n, p_x, p_y| dp_x dp_y$$
$$- e_0 \vec{\varepsilon} \bullet \frac{\vec{E}_{A0}}{2} \left[ e^{i(\omega_A t - k_A \hat{y} + \phi_A)} + e^{-i(\omega_A t - k_A \hat{y} + \phi_A)} \right] \quad (7)$$
$$- e_0 \vec{\varepsilon} \bullet \frac{\vec{E}_{B0}}{2} \left[ e^{i(\omega_B t + k_B \hat{y} + \phi_B)} + e^{-i(\omega_B t + k_B \hat{y} + \phi_B)} \right],$$

where $\vec{E}_{A0}$ and $\vec{E}_{B0}$ are vectors denoting the magnitude and polarization of the fields traveling in the +/- *y* directions respectively, $\vec{\varepsilon}$ is the position vector of the electron, and $e_0$ is the electron charge. Please refer to appendix B for the complete derivation of the wavefunction evolution in these fields. Simply the results are presented here.

If the atom begins completely in state $|1, \Psi_e(\vec{r},t)\rangle$ then after a time *T* of evolving in the above described fields, the result is:

$$|\Psi(\vec{r},t=T)\rangle = \cos\left(\frac{\Omega}{2}T\right)|1, \Psi_e(\vec{r},0)\rangle - ie^{i(\omega_B-\omega_A)T+i(\phi_B-\phi_A)}\sin\left(\frac{\Omega}{2}T\right)|3, \Psi_e(\vec{r},0)\rangle e^{-i(k_A+k_B)y},$$
(8)

where we have used the definitions given in the "formalism" section above. We see that for a π pulse ($T = \pi/\Omega$), Eq. (8) becomes:

$$|\Psi(x,y,t=\pi/\Omega)\rangle = -ie^{i(\omega_B-\omega_A)\frac{\pi}{\Omega}+i(\phi_B-\phi_A)}|3, \Psi_e(x,y,0)\rangle e^{-i(k_A+k_B)y},$$
(9)

while for a π/2 pulse ($T = \pi/(2\Omega)$), Eq. (8) yields:

$$|\Psi(x,y,t=\pi/(2\Omega))\rangle = \frac{1}{\sqrt{2}}|1, \Psi_e(x,y,0)\rangle$$
$$- ie^{i(\omega_B-\omega_A)\frac{\pi}{2\Omega}+i(\phi_B-\phi_A)}\frac{1}{\sqrt{2}}|3, \Psi_e(x,y,0)\rangle e^{-i(k_A+k_B)y}.$$
(10)

Similarly, if the atom begins completely in state $|3, \Psi_e(\vec{r},t)\rangle$, the wavefunction after a time *T* becomes:

$$|\Psi(x,y,t=T)\rangle = -ie^{i(\omega_A-\omega_B)T+i(\phi_A-\phi_B)}\sin\left(\frac{\Omega}{2}T\right)|1, \Psi_e(x,y,0)\rangle e^{i(k_A+k_B)y}$$
$$+ \cos\left(\frac{\Omega}{2}T\right)|3, \Psi_e(x,y,0)\rangle,$$
(11)

so that for a π pulse, Eq. (11) gives:

$$|\Psi(x,y,t=\pi/\Omega)\rangle = -ie^{i(\omega_A-\omega_B)\frac{\pi}{\Omega}+i(\phi_A-\phi_B)}|1,\Psi_e(x,y,0)\rangle e^{i(k_A+k_B)y}, \tag{12}$$

and for a π/2 pulse, Eq. (11) becomes:

$$|\Psi(x,y,t=\pi/(2\Omega))\rangle = -ie^{i(\omega_A-\omega_B)\frac{\pi}{2\Omega}+i(\phi_A-\phi_B)}\frac{1}{\sqrt{2}}|1,\Psi_e(x,y,0)\rangle e^{i(k_A+k_B)y}$$
$$+\frac{1}{\sqrt{2}}|3,\Psi_e(x,y,0)\rangle. \tag{13}$$

### III. D. State Evolution Through the Whole Interferometer

To see the effects of phase explicitly, we make use of the analysis that we have done for the state evolution of the wavepacket. Take our initial wavepacket $|\Psi\rangle$ to have initial conditions as discussed in the "formalism" section. At time $t=0$ the first π/2 pulse equally splits $|\Psi\rangle$ into two components $|\Psi_a\rangle$ and $|\Psi_b\rangle$ such that:

$$|\Psi_a\rangle = -ie^{i(\omega_B-\omega_A)\frac{\pi}{2\Omega}+i(\phi_{B1}-\phi_{A1})}\frac{1}{\sqrt{2}}|3,\Psi_e(x,y,0)\rangle e^{-i(k_A+k_B)y} \tag{14a}$$

$$|\Psi_b\rangle = \frac{1}{\sqrt{2}}|1,\Psi_e(x,y,0)\rangle, \tag{14b}$$

where we used Eq. (8). After a time $t=T_0$ of freespace (Eq. (6b)) and then a π pulse, Eqs. (8) and (11) yield:

$$|\Psi_a\rangle = -e^{i(\omega_A-\omega_B)\frac{\pi}{2\Omega}+i(\phi_{A2}-\phi_{A1}+\phi_{B1}-\phi_{B2})-i\omega_3 T_0}\frac{1}{\sqrt{2}}|1,\Psi_e(x,y-y_0,T_0)\rangle \tag{15a}$$

$$|\Psi_b\rangle = -ie^{i(\omega_B-\omega_A)\frac{\pi}{\Omega}+i(\phi_{B2}-\phi_{A2})-i\omega_1 T_0}\frac{1}{\sqrt{2}}|3,\Psi_e(x,y,T_0)\rangle e^{-i(k_A+k_B)y}. \tag{15b}$$

The $|\Psi_a\rangle$ component becomes shifted in space by $y_0$ due to the momentum it gained in the $+y$ direction from the π pulse. Now another zone of freespace for a time $T_0$ (Eq. (6b)) followed by the final π/2 pulse (using Eqs. (8) and (11)) forms:

$$|\Psi_a\rangle = -e^{i(\omega_A-\omega_B)\frac{\pi}{2\Omega}+i(\phi_{A2}-\phi_{A1}+\phi_{B1}-\phi_{B2})-i(\omega_1+\omega_3)T_0}\frac{1}{2}|1,\Psi_e(x,y-y_0,2T_0)\rangle$$
$$+ie^{i(\phi_{A2}-\phi_{A1}-\phi_{A3}+\phi_{B1}-\phi_{B2}+\phi_{B3})-i(\omega_1+\omega_3)T_0}\frac{1}{2}|3,\Psi_e(x,y-y_0,2T_0)\rangle e^{-i(k_A+k_B)y} \quad (16a)$$

and

$$|\Psi_b\rangle = -e^{i(\omega_B-\omega_A)\frac{\pi}{2\Omega}+i(\phi_{B2}-\phi_{B3}-\phi_{A2}+\phi_{A3})-i(\omega_1+\omega_3)T_0}\frac{1}{2}|1,\Psi_e(x,y-y_0,2T_0)\rangle$$
$$-ie^{i(\omega_B-\omega_A)\frac{\pi}{\Omega}+i(\phi_{B2}-\phi_{A2})-i(\omega_1+\omega_3)T_0}\frac{1}{2}|3,\Psi_e(x,y-y_0,2T_0)\rangle e^{-i(k_A+k_B)y}. \quad (16b)$$

Now the $|\Psi_b\rangle$ component is spatially aligned with the $|\Psi_a\rangle$ component. However, another split occurs because both of these components are partially in internal state $|3\rangle$. After some further time $T_1$ in freespace, state $|3\rangle$ has drifted further in the $+y$ direction. The substrate can now intercept the two internal states of the total wavefunction in separate locations. We write the state $|1\rangle$ wavefunction as:

$$|\Psi_1\rangle = -\frac{1}{2}\left(e^{i(\omega_B-\omega_A)\frac{\pi}{2\Omega}+i(\phi_{B2}-\phi_{B3}-\phi_{A2}+\phi_{A3})} + e^{i(\omega_A-\omega_B)\frac{\pi}{2\Omega}+i(\phi_{A2}-\phi_{A1}+\phi_{B1}-\phi_{B2})}\right)$$
$$\times|1,\Psi_e(x,y-y_0,2T_0+T_1)\rangle e^{-i(\omega_1+\omega_3)T_0} \quad (17a)$$

and the state $|3\rangle$ wavefunction as:

$$|\Psi_3\rangle = i\frac{1}{2}\left(e^{i(\phi_{A2}-\phi_{A1}-\phi_{A3}+\phi_{B1}-\phi_{B2}+\phi_{B3})} - e^{i(\omega_B-\omega_A)\frac{\pi}{\Omega}+i(\phi_{B2}-\phi_{A2})}\right)$$
$$\times|3,\Psi_e(x,y-y_0-y_1,2T_0+T_1)\rangle e^{-i(k_A+k_B)y-i(\omega_1+\omega_3)T_0} \quad (17b)$$

These have populations:

$$\langle \Psi_1 | \Psi_1 \rangle = \frac{1}{2}(1 + \cos(\phi_0)), \qquad \langle \Psi_3 | \Psi_3 \rangle = \frac{1}{2}(1 - \cos(\phi_0)), \qquad (18)$$

where $\phi_0 = \frac{\pi}{\Omega}(\omega_A - \omega_B) - \phi_{A1} + \phi_{B1} + 2\phi_{A2} - 2\phi_{B2} - \phi_{A3} + \phi_{B3}$. We see that the state populations are functions of the phase differences of the laser fields. Since we can choose these phase differences arbitrarily, we can populate the states arbitrarily. If we choose the phases, for example, such that $\phi_0$ is some multiple of $2\pi$, then the wavepacket population will end up entirely in internal state $|1\rangle$.

## IV. Arbitrary Image Formation

If, however, between the $\pi$ pulse and second $\pi/2$ pulse we apply a spatially varying phase shift $\phi_P(\vec{r})$ to $|\Psi_a\rangle$, but keep $\phi_0$ as a multiple of $2\pi$, then the populations in Eqs. (18) become instead:

$$\langle \Psi_1 | \Psi_1 \rangle = \frac{1}{2}(1 + \cos(\phi_P(\vec{r}))), \qquad \langle \Psi_3 | \Psi_3 \rangle = \frac{1}{2}(1 - \cos(\phi_P(\vec{r}))). \qquad (19)$$

Therefore, if we let $\phi_P(\vec{r}) = \arccos(P(\vec{r}))$, where $P(\vec{r})$ is an arbitrary pattern normalized to 1, the state $|1\rangle$ population will be:

$$\langle \Psi_1 | \Psi_1 \rangle = \frac{1}{2}(1 + P(\vec{r})). \qquad (20)$$

If the substrate at (8) in Fig. 1 intercepts just this state, the population distribution will be in the form of the arbitrary image. Over time, depositions on the substrate will follow the population distribution, and thereby physically form the image on the substrate.

### IV. A. Imparting an Arbitrary, Spatially Varying Phase Shift for Arbitrary Image Formation

We now review how to do such phase imprinting [24, 25] to a single wavepacket using the ac-stark effect.

First, consider the SE for the wavepacket expressed in position space:

$$i\hbar \frac{\partial |\Psi(\vec{r},t)\rangle}{\partial t} = \frac{-\hbar^2}{2m}\nabla^2|\Psi(\vec{r},t)\rangle + V(\vec{r})|\Psi(\vec{r},t)\rangle. \tag{21}$$

If we consider a very short interaction time $\tau$ with the potential $V(\vec{r})$, we find:

$$i\hbar \frac{\partial |\Psi(\vec{r},t+\tau)\rangle}{\partial t} \approx V(\vec{r})|\Psi(\vec{r},t+\tau)\rangle$$

$$\Rightarrow |\Psi(\vec{r},t+\tau)\rangle \approx |\Psi(\vec{r},t)\rangle e^{-\frac{i}{\hbar}V(\vec{r})\tau} \tag{22}$$

Thus, we see that an arbitrary phase shift $\phi_P(\vec{r})$ is imparted on the wavepacket in position space by choosing $V(\vec{r}) = (\hbar/\tau)\phi_P(\vec{r})$. Although this would give the negative of the desired phase, it makes no difference because it is the cosine of the phase that gives the interference pattern.

In order to create the arbitrary potential needed to impart the arbitrary phase shift, we use the ac-*Stark effect* (*light shift*). As illustrated in Fig. 1 at 4b, 5b, and 6b, the atom will be in the internal state $|1\rangle$. If exposed to a highly detuned laser field that only excites the $|1\rangle \rightarrow |2\rangle$ transition, the eigenstates become perturbed such that their energies shift in proportion to the intensity of the laser field. A spatially varying intensity will yield a spatially varying potential energy. Specifically, in the limit that $g/\delta \rightarrow 0$, where $g$ is proportional to the square root of the laser intensity and $\delta$ is the detuning, it is found that the energy of the ground state is approximately $\hbar g^2/(4\delta)$. To impart the pattern phase, then, we subject the atomic wavepacket at 4b, 5b, and 6b in Fig. 1 to a laser field that has an intensity variation in the *x-y* plane such that:

$$\begin{aligned} g^2(\vec{r}) &= (4\delta/\tau)\phi_P(\vec{r}) \\ &= (4\delta/\tau)\arccos(P(\vec{r})), \end{aligned} \tag{23}$$

where $P(\vec{r})$ is the arbitrary pattern normalized to 1 and $\Delta t$ is the interaction time.

### IV. B. The Need for a Lens System

The need for a lens system for the atomic wavepacket arises due to two separate considerations. First, there is a need for expanding and focusing the wavepacket in order to shrink down the phase pattern imparted at (5b) in Fig. 1. We have shown above how the phase pattern is imparted using an intensity variation on an impinging light pulse. However, due to the diffraction limit of light, the scale limit of this variation will be on the order of 100nm. This will cause the interference at (8) to occur on that scale. To reach a smaller scale, we require a lens system that allows expansion and focusing of the wavepacket to occur in the transverse plane. Using such a system, we could, for example, expand the wavepacket by two orders of magnitude prior to (5b), impart the phase pattern at (5b), and then focus it back to its original size by the time it reaches (8). The interference would then occur on the scale of 1 nm.

The second consideration which must be made is that an arbitrary phase shift $\phi(x,y)$ introduced at (5b), if it has any variation at all in the transverse plane, will cause the wavepacket traveling along that arm of the AI to alter its momentum state. Any freespace evolution after this point will make the wavepacket distort or go off trajectrory, causing a noisy interference or even eliminating interference at (8) all together.

Our lens system, then, must accomplish two objectives simultaneously: 1) allow for an expansion and focusing of the wavepacket to occur and 2) have the wavepacket properly aligned and undistorted when it reaches (8). To do this, we employ techniques similar to those developed in classical Fourier optics [3]. First we develop a diffraction theory for the 2D quantum mechanical wavepacket, then we use the theory to setup a lens system that performs spatial Fourier transforms on the wavepacket in order to achieve the two above stated objectives.

### IV. C. Development of the Quantum Mechanical Wavefunction Diffraction Theory

Consider the 2D SE in freespace:

$$i\hbar \frac{\partial |\Psi(\vec{r},t)\rangle}{\partial t} = \frac{-\hbar^2}{2m}\left(\frac{\partial^2}{\partial x^2} + \frac{\partial^2}{\partial y^2}\right)|\Psi(\vec{r},t)\rangle. \tag{24}$$

By inspection, we see that it is linear and shift independent. If we can then find the impulse response of this "system" and convolve it with an arbitrary input, we can get an exact analytical expression for the output. To proceed, we first try to find the transfer function of the system.

Using the method of separation of variables, it is readily shown that all solutions of the system (the 2D SE in free space) can be expressed as linear superpositions of the following function:

$$|\Psi(\vec{r},t)\rangle = A e^{i(\vec{k}\cdot\vec{r} - \frac{\hbar}{2m}|\vec{k}|^2 t)}, \tag{25}$$

where $A$ is some constant and $\vec{k} = k_x \hat{i} + k_y \hat{j}$ can take on any values. Now let us take some arbitrary input to our system at time $t=0$ and express it in terms of its Fourier components:

$$|\Psi_{in}(\vec{r})\rangle = \frac{1}{2\pi}\int \Phi_{in}(\vec{k}) e^{i\vec{k}\cdot\vec{r}} d\vec{k}. \tag{26}$$

We can then evolve each Fourier component for a time $T$ by using Eq. (25) to get the output:

$$|\Psi_{out}(\vec{r})\rangle = \frac{1}{2\pi} \int |\Phi_{in}(\vec{k})\rangle e^{i(\vec{k}\bullet\vec{r} - \frac{\hbar}{2m}|\vec{k}|^2 T)} d\vec{k}$$
$$= \frac{1}{2\pi} \int \left(|\Phi_{in}(\vec{k})\rangle e^{-i\frac{\hbar}{2m}|\vec{k}|^2 T}\right) e^{i\vec{k}\bullet\vec{r}} d\vec{k} \quad (27)$$
$$= \frac{1}{2\pi} \int |\Phi_{out}(\vec{k})\rangle e^{i\vec{k}\bullet\vec{r}} d\vec{k}.$$

It follows that:

$$|\Phi_{out}(\vec{k})\rangle = |\Phi_{in}(\vec{k})\rangle e^{-i\frac{\hbar}{2m}|\vec{k}|^2 T}. \quad (28)$$

Our transfer function, then, for a free space system of time duration $T$ is:

$$H(\vec{k}) = e^{-i\frac{\hbar}{2m}|\vec{k}|^2 T}. \quad (29)$$

After taking the inverse Fourier transform, we find the impulse response to be:

$$h(\vec{r}) = -i\left(\frac{m}{\hbar T}\right) e^{i\left(\frac{m}{2\hbar T}\right)|\vec{r}|^2}. \quad (30)$$

Finally, convolving this with some input to the system at time $t=0$, $|\Psi_{in}(\vec{r})\rangle$, gives the output at time $t=T$, $|\Psi_{out}(\vec{r})\rangle$, to be:

$$|\Psi_{out}(\vec{r})\rangle = -i\left(\frac{1}{2\pi}\frac{m}{\hbar T}\right) e^{i\frac{m}{2\hbar T}|\vec{r}|^2} \int |\Psi_{in}(\vec{r}')\rangle e^{i\frac{m}{2\hbar T}|\vec{r}'|^2} e^{-i\frac{m}{\hbar T}\vec{r}\bullet\vec{r}'} d\vec{r}'. \quad (31)$$

This expression is analogous to the *Fresnel diffraction integral* [3] from classical optics.

**IV. D.  A Fourier Transform Lens Scheme**

Consider now the following:

1) Take as input some wavefunction $|\Psi(\vec{r})\rangle$, and use the light shift to apply a "lens" (in much the same was as we show above how to apply the arbitrary pattern phase) such that it becomes:

$$|\Psi(\vec{r})\rangle e^{-i\frac{m}{2\hbar T}|\vec{r}|^2}$$

2) Pass it through the free space system for a time $T$ using the above derived integral to get:

$$-i\left(\frac{1}{2\pi}\frac{m}{\hbar T}\right) e^{i\frac{m}{2\hbar T}|\vec{r}|^2} \int |\Psi(\vec{r}')\rangle e^{-i\frac{m}{\hbar T}\vec{r}\cdot\vec{r}'} d\vec{r}'$$

3) Now use the light shift again to create another "lens" where the phase shift is $e^{-i(\frac{m}{2\hbar T}|\vec{r}|^2 - \frac{\pi}{2})}$ so that we are left with:

$$\left(\frac{1}{2\pi}\frac{m}{\hbar T}\right) \int |\Psi(\vec{r}')\rangle e^{-i\frac{m}{\hbar T}\vec{r}\cdot\vec{r}'} d\vec{r}'.$$

We see that this is simply a scaled version of the Fourier transform of the input.  This lens system, then, is such that:

$$|\Psi_{out}(\vec{r})\rangle = \left(\frac{1}{2\pi}\frac{m}{\hbar T}\right) \left|\Phi_{in}\left(\frac{m}{\hbar T}\vec{r}\right)\right\rangle, \qquad (32)$$

where $|\Phi_{in}\rangle = F.T.\{|\Psi_{in}\rangle\}$.

**IV. E. Using the F.T. lens scheme to Create a Distortion Free Expansion and Focusing System for applying the Pattern Phase**

In order to achieve our desired goals of doing expansion/focusing and preventing distortion, we propose the system illustrated in Fig. 2a.  We first input our Gaussian wavepacket into a F.T. scheme with a characteristic time parameter $T = T_A$. We will then get the Fourier transform of the input (also a Gaussian) scaled by $m/(\hbar T_A)$. Then, we give the wavepacket a phase shift that corresponds to the desired interference pattern (pattern phase) and put it through another F.T. scheme with the same time parameter $T_A$. The wavefunction is now the convolution of the original input with the pattern phase. Finally, a third F.T. scheme is used with $T = T_B$ so that the output is the same as the wavefunction just before the second F.T. scheme, but is now reflected about the origin and scaled by $m/(\hbar T_B)$ instead of $m/(\hbar T_A)$. The pattern phase, therefore, has been scaled down by a factor of $T_A/T_B$. Since both $T_A$ and $T_B$ can be chosen arbitrarily, we can, in principle, scale down the pattern phase by orders of magnitude. If, for example, the pattern phase is first imparted on a scale of ~100 nm, we can choose $T_A/T_B$ to be 100 so that at the output of our lens system, it is on a scale of ~1 nm. By scaling down the pattern phase, we can scale down the interference pattern at (8) in Fig. 1.

Within the context of the interferometer, our lens system is placed at (4b), (5b), and (6b) in Fig. 1. Now, since the system provides us with the desired output immediately in time after the final lens (lens 3b in Fig. 2a), this final lens, the final π/2 pulse, and the substrate (6) all need to be adjacent. If they are not, the wavepacket will undergo extra freespace evolution and may distort. However, such a geometry is difficult to achieve experimentally so we propose a modification to the lens system (Fig. 2b). Specifically, we can move the lens 3b in Fig. 2a to occur immediately before lens 2a, as long as we rescale it to account for the different wavepacket size at that location. We call the rescaled version $3b'$, which is the same as 3b except for the parameter $T_A$ in place of $T_B$. We can then place the substrate at (8) in Fig. 2a to be where the lens 3b previously was; that is, a time $T_B$ away from lens 3a. The final π/2 pulse can occur anywhere between lens 3a and the substrate, as long as it is far enough away from the substrate to allow sufficient time for the state $|3\rangle$ component to separate from the state $|1\rangle$ component. To avoid disturbing the requisite symmetry of the AI, we accomplish this by choosing $T_B$ to be sufficiently large while leaving the final π/2 pulse itself in its original location. This geometry will allow the substrate to intercept the state $|1\rangle$ component exclusively and at precisely the right moment such that it does not undergo too little or too much freespace evolution without having any of the final π/2 pulse, final lens, or substrate adjacent. Finally, we can simplify the lens system's implementation if we combine the lenses that are adjacent. Lenses 1b, 2a, $3b'$, and $\phi_P(\bar{r})$ can be combined into lens α; lenses 2b and 3a can be combined into lens β. Explicitly, lens α has phase shift:

$$\phi_\alpha(\bar{r}) = -\left(\frac{3m}{2\hbar T_A}\right)|\bar{r}|^2 + \pi - \phi_P(\bar{r}), \qquad (33)$$

and lens β has phase shift:

$$\phi_\beta(\bar{r}) = -\left(\frac{m}{2\hbar T_A} + \frac{m}{2\hbar T_B}\right)|\bar{r}|^2 + \frac{\pi}{2}. \qquad (34)$$

Fig. 2c shows the implementation of the lens system within the context of the whole AI.

A cause for concern may arise in the fact that with the lens system in place, the part of the wavepacket that travels along the arm without the lens will be interfering not with a phase modified version of itself, but with a phase modified Fourier transform of itself. That is, the output of the lens system is a phase modified Fourier transform of its input. As such, the effective width of the wavepacket coming from the lens system may be significantly larger than the effective width of that coming from the arm without lenses, thus causing a truncation of the pattern formation around the edges. This problem is addressed by selecting $T_B$ such that the wavepacket from the lens system is scaled to have an effective width equivalent to or smaller than the wavepacket from the other arm. Also, because of the Fourier transform, the wavepacket coming from the lens system, even without an added pattern phase, may have a different phase signature than the wavepacket coming from the other arm. Regarding this issue, our numerical experiments show that after freespace propagation for a time on the order of the timescale determined as practical (see section on practical considerations), the phase difference between the original wavepacket and its Fourier transform is very small over the span of the effective width of the wavepacket. Thus, the effect of this phase noise on the interference pattern is negligible.

## V. Some Practical Considerations

### V. A. Wavepacket Behavior

The behavior of the wavepacket primarily has implications for the time and wavepacket effective width parameters of the lithography scheme. As mentioned earlier, the scale limit of the intensity variation that creates the pattern phase when it is first applied is $\sim 10^7$ meters. The lens system then further reduces the scale of the pattern phase by a factor of $T_A/T_B$. To achieve lithography features on the scale of ~1nm, this ratio needs to be ~100. However, we must also take into consideration the extent of the entire intensity variation. In other words, referring to Fig. 2c, the effective width of the wavepacket at lens α must be large enough to accommodate the entire pattern on the light pulse bearing the phase pattern information. We assume that this dimension will be on the order of a millimeter. We know that the wavepacket at lens α is a scaled Fourier transform of the wavepacket immediately before lens 1a, so that its effective width at lens α is $\dfrac{\hbar T_A}{m\sigma_{in}}$. This must be on the order of $10^{-3}$. Also, another way in which the time parameters are restricted is by the total amount of time that the atom spends in the AI. Even with a magnetic field slowing the atom's fall, it is not practical to have the atom spend a large amount of time in the chamber. We therefore impose the restriction that the atom spend no more time than a minute or two in the chamber. Explicitly, this translates to: $T_A \leq \sim 10^1$.

Now, as shown earlier, it is the state $|1\rangle$ component in our scheme that will form the desired interference pattern. The substrate must therefore intercept this component exclusive of the state $|3\rangle$ component. Fortunately, the state $|3\rangle$ component will have an additional velocity in the *y* direction due to photon recoil so that the two states will separate if given enough time. Also recall that each wavepacket state after the final π/2 pulse is composed of two elements, one that went through the lens system and one that did not, such that the elements that traveled along the arm without the lens system will have larger effective widths (since the output of the lens system is smaller than its input). The two states will be sufficiently separated, then, when the state $|3\rangle$ component has traveled far enough in the $+y$ direction after the final $\pi/2$ pulse such that there is no overlap of the larger effective widths. Since we know that photon recoil gives the state $|3\rangle$ component an additional momentum of $2\hbar k$ in the $+y$ direction, we have: $mv = 2\hbar k$. Also, it can be shown that the effective width of a wavepacket after passing through freespace for a time $T$ is $\sigma\sqrt{1+(T/\tau)}$, where $\tau = m\sigma/\hbar$ and $\sigma$ is the original effective width. Therefore, for sufficient spatial separation of the states (assuming that the time between the final $\pi/2$ pulse and the substrate is on the order of $T_B$) we need: $v*T_B \geq \sim \sigma_{in}\sqrt{1+(T_B/\tau)}$.

To summarize, our restrictions are:

$$\boxed{T_A \leq \sim 10^1} \quad \boxed{T_B \leq \sim 10^{-2} T_A} \quad \boxed{\frac{\hbar T_A}{m\sigma_{in}} \geq \sim 10^{-3}} \quad \boxed{\frac{2\hbar k}{m} T_B \geq \sim \sigma_{in}\sqrt{1+\left(\frac{\hbar T_B}{m\sigma_{in}}\right)}}.$$

After using some simple algebra, we find that the first three restrictions are satisfied if we apply the following:

$$\boxed{\sigma_{in} \leq 10^{-5}} \quad \boxed{\sim 10^6 \sigma_{in} \leq T_A \leq \sim 10^1} \quad \boxed{T_B \leq \sim 10^{-2} T_A}.$$

We can, for example, choose: $\sigma_{in} = 10^{-5}$, $T_A \sim 10^1$, $T_B \sim 10^{-1}$. A simple check shows that these choices also satisfy the fourth restriction.

Finally, since our proposed lithography scheme involves the use of a single atom at a time, it entails the drawback of being very slow. To make this type of lithography truly practical, a Bose-Einstein condensate [4, 5] would have to be used instead of a single atomic wavepacket.

## V. B. $^{87}$Rubidium Transition Scheme

For practical implementation of our three-level atom, we use the D1 transitions in $^{87}$Rb [40]. Fig. 3 illustrates. One of the restrictions is that, in order to be able to neglect spontaneous emission, we need for each single transition:

$$\left(\frac{g_0}{\delta}\right)^2 *\Gamma*\tau \ll 1, \qquad (35)$$

where $g_0$ is the Rabi frequency, $\delta$ is the detuning, $\Gamma$ is the decay rate, and $\tau$ is the interaction time. Both the Raman pulse scheme and the light shift scheme also require:

$$g_0 \ll \delta. \qquad (36)$$

Each of the transitions that we have chosen are the strongest ones from their group, so we assume them both to have a saturation intensity of about $3 \, mW/cm^2$. We have the following relation:

$$g_{0,max}^2 = \left(\frac{I_{max}}{I_{sat}}\right)\Gamma^2. \tag{37}$$

If we assume $I_{max} = 2 \, W/mm^2$ and $\Gamma = 3.33*10^7 \, sec^{-1}$, we find that $g_{0,max} \approx 8.6*10^9 \, Hz$.

Now, the $\sigma^+$ polarized light that we have chosen excites not only the desired transition from $|1\rangle$ to $|2\rangle$, but also a transition from $|2\rangle$ to a different metastable state which we do not wish to populate. Fortunately, this second transition is about 6.8 GHz less than the first one, and its coupling strength is about 4 times smaller. The $\pi$ polarized light that we have chosen for the $|2\rangle$ to $|3\rangle$ transition also excites undesired transitions by linking state $|1\rangle$ to excited states other than the one we have chosen for $|2\rangle$. However, the undesired transitions that are excited in this case are also over 6 GHz larger than the desired transition. We can therefore neglect the unwanted transitions by making sure that our $\sigma^+$ polarized and $\pi$ polarized pulses are detuned from their appropriate transitions by no more than a few hundred MHz, thereby assuring that the detuning for the unwanted transitions is at least a factor of 10 greater than the detuning for the desired transitions. We choose our detuning to be 680 MHz.

In order to satisfy the constraint that the Rabi frequency be much less than the detuning, we choose $g_0 = 68$ MHz. This is well below the maximum limit calculated above.

As far as the interaction time for the π/2 and π pulse scheme, it is the Raman Rabi frequency that is of interest:

$$\Omega = \frac{g_0^2}{2\delta}. \tag{38}$$

Using this in Eq. (35), we get:

$$2\frac{\Omega}{\delta}*\Gamma*\tau << 1$$

$$\Rightarrow \quad \Omega\tau << \frac{\delta}{2\Gamma} \tag{39}$$

Plugging in the chosen value for δ and the typical value of 33.33 MHz for Γ, we find that $\Omega\tau << 10.2$. We can satisfy this restraint by choosing $\Omega\tau = \pi$ for the π pulse and half as much for the π/2 pulse, giving a pulse duration of $\tau = \pi/\Omega \approx 924$ ns for a π pulse and $\tau \approx 462$ ns for a π/2 pulse.

For the light shift we use the same π polarized excitation of state $|3\rangle$ as above. The time constraint in this case is:

$$\frac{g_0^2}{4\delta}\tau = 2\pi. \tag{40}$$

This gives an interaction time of $\tau \approx 3.7\,\mu\text{sec}$. Ideally, the light shift pulse will only interact with the wavepacket in state $|3\rangle$. This may actually be possible if we choose $T_A$ to be large enough such that the two states gain enough of a transverse separation. If, as by example above, we choose $T_A \sim 10^1$, then the separation between the two states will be on the order of a centimeter and there will be virtually no overlap between the two components of the wavepacket in the separate arms. The light pulse could then simply intercept only state $|3\rangle$. If, however, the situation is such that the states are overlapping, then state $|1\rangle$ will also see the light shift, but it will be about a factor of 10 less because of the detuning being approximately 10 times larger for it than for the state $|3\rangle$ transition.

## VI. Numerical Experiments

The numerical implementation of our lithography scheme was done in Matlab$^{\text{TM}}$ by distributing the wavepackets across finite meshes and then evolving them according to the Schrodinger equation. This evolution was done in both position and momentum space according to expediency. To go between the two domains, we used two-dimensional Fourier Transform and Inverse Fourier Transform algorithms.

The initial wavepacket was taken in momentum space and completely in internal state $|1\rangle$. Specifically, the wavepacket was given by the Fourier Transform of Eq. (1):

$$\left|\Phi_e(\vec{k},t=0)\right\rangle = \sqrt{\frac{\sigma}{\sqrt{\pi}}}\exp\left(\frac{-|\vec{k}|^2\sigma^2}{2}\right). \tag{41}$$

The evolution of the wavepackets in the $\pi$ and $\pi/2$ pulses was done in momentum space in order to be able to account for the different detunings that result for each momentum component due to the Doppler shift. Specifically, we numerically solved Eq. (B15) for the different components of the k-space wavepacket mesh, then applied the inverse of the transformation matrix given by Eq. (B9) to go to the original basis.

Outside of the lens system, the free space evolution of the wavepackets was also done in momentum space. This was achieved easily by using Eqs. (A4). Within the lens system, however, it was more computationally efficient to use Eq. (31) for the freespace evolution because of the need to apply the lenses in position space. The results of using Eq. (31) were initially cross-checked with the results of using Eqs. (A4) and were found to agree.

Fig. 4a-b demonstrate the formation of an arbitrary pattern by interference of the state $|1\rangle$ wavepackets at the output of the interferometer. Both figures were the result of applying the same arbitrary pattern phase, but Fig. 4a was formed without any shrinking implemented (i.e. $T_A=T_B$). Fig. 4b, however, demonstrates the shrinking ability of the lens system by yielding a version of Fig. 4a that is scaled by a factor of two ($T_A/T_B = 2$). The length scales are in arbitrary units due to the use of naturalized units for the sake of computational viability.

## VII. Suggestions for Extension to BEC

As mentioned above, in order to make the lithography scheme truly practical, a Bose-Einstein condensate is required in place of the single atom. Indeed, the self-interference of a BEC has already been demonstrated [41, 42]. The difficulty in using the BEC for controlled imaging, however, arises from the nonlinear term in the Gross-Pitaevskii equation (GPE). Our lens system, for example, would not be valid as it was developed from the linear SE.

One approach to getting around this problem is to try to eliminate the nonlinear term in the GPE. Specifically, the GPE for the BEC takes the form

$$i\hbar \frac{\partial \Psi}{\partial t} = \left( \frac{-\hbar^2}{2m} \nabla^2 + V + U_0 |\Psi|^2 \right) \Psi, \qquad (42)$$

where the nonlinear term coefficient is $U_0 = 4\pi\hbar^2 a/m$ and $a$ is the scattering length for the atom. It has been demonstrated for $^{87}$Rb that the scattering length can be tuned over a broad range by exposing the BEC to magnetic fields of varying strength near Feshbach resonances [43, 44]. The relationship between the scattering length and the applied magnetic field $B$ when near a Feshbach resonance can be written as

$$a = a_{bg}\left(1 - \frac{\Delta}{B - B_{peak}}\right), \qquad (43)$$

where $a_{bg}$ is the background scattering length, $B_{peak}$ is the resonance position, and $\Delta = B_{zero} - B_{peak}$. Setting $B = B_{zero}$ would therefore set the scattering length to zero and eliminate the nonlinear term in the GPE. While the atom-atom interaction may not be completely eliminated in reality due to the fluctuation in density that we wish to effect through the lens system, it is worth investigating if it could be made to be negligible over an acceptable range. We could then use our previously developed lens system to perform the imaging and thereby interfere thousands or millions of atoms simultaneously.

This work was supported by DARPA grant # F30602-01-2-0546 under the QUIST program, ARO grant # DAAD19-001-0177 under the MURI program, and NRO grant # NRO-000-00-C-0158.

# Appendix A
**State Evolution in Free Space**

In free space, the Hamiltonian can be expressed in the momentum domain as:

$$H = \iint \sum_{n=1}^{3} \left( \frac{p_x^2 + p_y^2}{2m} + \hbar\omega_n \right) |n, p_x, p_y\rangle\langle n, p_x, p_y| \, dp_x dp_y. \tag{A1}$$

Where $\omega_n$ is the frequency corresponding to the eigenenergy of internal state $|n\rangle$. For a single momentum component ($p_x = p_{x0}$ and $p_y = p_{y0}$), the Hamiltonian for the total evolution in momentum space is given by:

$$H = \begin{bmatrix} \frac{p_{x0}^2 + p_{y0}^2}{2m} + \hbar\omega_1 & 0 & 0 \\ 0 & \frac{p_{x0}^2 + p_{y0}^2}{2m} + \hbar\omega_2 & 0 \\ 0 & 0 & \frac{p_{x0}^2 + p_{y0}^2}{2m} + \hbar\omega_3 \end{bmatrix}. \tag{A2}$$

Using this in the SE, we get the equations of the amplitude evolution in momentum space:

$$\dot{C}_1(p_{x0}, p_{y0}, t) = -\frac{i}{\hbar}\left(\frac{p_{x0}^2 + p_{y0}^2}{2m} + \hbar\omega_1\right)C_1(p_{x0}, p_{y0}, t),$$

$$\dot{C}_2(p_{x0}, p_{y0}, t) = -\frac{i}{\hbar}\left(\frac{p_{x0}^2 + p_{y0}^2}{2m} + \hbar\omega_2\right)C_2(p_{x0}, p_{y0}, t), \tag{A3}$$

$$\dot{C}_3(p_{x0}, p_{y0}, t) = -\frac{i}{\hbar}\left(\frac{p_{x0}^2 + p_{y0}^2}{2m} + \hbar\omega_3\right)C_3(p_{x0}, p_{y0}, t).$$

These yield the solutions:

$$C_1(p_{x0}, p_{y0}, t) = C_1(p_{x0}, p_{y0}, 0) e^{-i\left(\frac{p_{x0}^2 + p_{y0}^2}{2m\hbar} + \omega_1\right)t},$$

$$C_2(p_{x0}, p_{y0}, t) = C_2(p_{x0}, p_{y0}, 0) e^{-i\left(\frac{p_{x0}^2 + p_{y0}^2}{2m\hbar} + \omega_2\right)t}, \qquad (A4)$$

$$C_3(p_{x0}, p_{y0}, t) = C_3(p_{x0}, p_{y0}, 0) e^{-i\left(\frac{p_{x0}^2 + p_{y0}^2}{2m\hbar} + \omega_3\right)t}.$$

We see that if the wavefunction is known at time $t = 0$, then after a duration of time $T$ in freespace, the wavefunction becomes:

$$|\Psi(\vec{r}, t = T)\rangle = \frac{1}{2\pi} \iint \Bigg[ C_1(p_x, p_y, 0) e^{-i\left(\frac{p_x^2 + p_y^2}{2m\hbar} + \omega_1\right)T} |1, p_x, p_y\rangle$$
$$+ C_2(p_x, p_y, 0) e^{-i\left(\frac{p_x^2 + p_y^2}{2m\hbar} + \omega_2\right)T} |2, p_x, p_y\rangle$$
$$+ C_3(p_x, p_y, 0) e^{-i\left(\frac{p_x^2 + p_y^2}{2m\hbar} + \omega_3\right)T} |3, p_x, p_y\rangle \Bigg] dp_x dp_y. \qquad (A5)$$

We can also write it as:

$$|\Psi(\vec{r}, t = T)\rangle = e^{-i\omega_1 T} c_1(0) |1, \Psi_e(\vec{r}, T)\rangle + e^{-i\omega_2 T} c_2(0) |2, \Psi_e(\vec{r}, T)\rangle$$
$$+ e^{-i\omega_3 T} c_3(0) |3, \Psi_e(\vec{r}, T)\rangle \qquad (A6)$$

# Appendix B
### State Evolution in π and π/2 pulse laser fields

The electromagnetic fields encountered by the atom at points 2, 3, and 7 in Fig. 1 that act as the π and π/2 pulses are each formed by two lasers that are counter-propagating in the y-z plane parallel to the y axis. We will refer to the laser propagating in the +y direction as $\vec{E}_A$, and the one propagating in the – *y* direction as $\vec{E}_B$. In deriving the equations of motion under this excitation, we make the following assumptions: (1) the laser fields can be treated semi-classically [45], (2) the intensity profiles of the laser fields forming the π and π/2 pulses remain constant over the extent of the atomic wavepacket, (3) the wavelengths of the lasers are significantly larger than the separation distance between the nucleus and electron of the atom, (4) $\vec{E}_A$ excites only the $|1\rangle \leftrightarrow |2\rangle$ transition and $\vec{E}_B$ only the $|3\rangle \leftrightarrow |2\rangle$ transition, (5) $\vec{E}_A$ and $\vec{E}_B$ are far detuned from the transitions that they excite, and (6) $\vec{E}_A$ and $\vec{E}_B$ are of the same intensity.

Using assumptions 1) and 2), we write the laser fields as:

$$\begin{aligned}\vec{E}_A &= \vec{E}_{A0} \cos(\omega_A t - k_A \hat{y} + \phi_A) \\ &= \frac{\vec{E}_{A0}}{2} \left[ e^{i(\omega_A t - k_A \hat{y} + \phi_A)} + e^{-i(\omega_A t - k_A \hat{y} + \phi_A)} \right]\end{aligned} \tag{B1}$$

and

$$\begin{aligned}\vec{E}_B &= \vec{E}_{B0} \cos(\omega_B t + k_B \hat{y} + \phi_B) \\ &= \frac{\vec{E}_{B0}}{2} \left[ e^{i(\omega_B t + k_B \hat{y} + \phi_B)} + e^{-i(\omega_B t + k_B \hat{y} + \phi_B)} \right]\end{aligned} \tag{B2}$$

where $\vec{E}_{A0}$ and $\vec{E}_{B0}$ are vectors denoting the magnitude and polarization of their respective fields. Keeping in mind that our wavefunction is expressed in the momentum domain, we take position as an operator.

The Hamiltonian here is expressed as the sum of two parts: $H = H_0 + H_1$. The first part corresponds to the non-interaction energy:

$$H_0 = \iint \sum_{n=1}^{3} \left( \frac{p_x^2 + p_y^2}{2m} + \hbar \omega_n \right) |n, p_x, p_y\rangle \langle n, p_x, p_y | dp_x dp_y . \tag{B3}$$

The second part accounts for the interaction energy, for which we use assumption (3) from above to make the electric dipole approximation and get:

$$H_1 = -e_0 \bar{\varepsilon} \bullet \frac{\bar{E}_{A0}}{2} \left[ e^{i(\omega_A t - k_A \hat{y} + \phi_A)} + e^{-i(\omega_A t - k_A \hat{y} + \phi_A)} \right]$$
$$-e_0 \bar{\varepsilon} \bullet \frac{\bar{E}_{B0}}{2} \left[ e^{i(\omega_B t + k_B \hat{y} + \phi_B)} + e^{-i(\omega_B t + k_B \hat{y} + \phi_B)} \right],$$
(B4)

where $\bar{\varepsilon}$ is the position vector of the electron, and $e_0$ is the electron charge. Now, seeing that expressions of the form $\langle n|\bar{\varepsilon} \bullet \bar{E}_{A0}|n\rangle$ and $\langle n|\bar{\varepsilon} \bullet \bar{E}_{B0}|n\rangle$ are zero, and using assumption (4), we can express Eq. (B4) as:

$$H_1 = \iint \left[ \frac{\hbar g_A}{2} \begin{pmatrix} |1,p_x,p_y\rangle\langle 2,p_x,p_y| \\ +|2,p_x,p_y\rangle\langle 1,p_x,p_y| \end{pmatrix} \left[ e^{i(\omega_A t - k_A \hat{y} + \phi_A)} + e^{-i(\omega_A t - k_A \hat{y} + \phi_A)} \right] \right.$$
$$\left. + \frac{\hbar g_B}{2} \begin{pmatrix} |3,p_x,p_y\rangle\langle 2,p_x,p_y| \\ +|2,p_x,p_y\rangle\langle 3,p_x,p_y| \end{pmatrix} \left[ e^{i(\omega_B t + k_B \hat{y} + \phi_B)} + e^{-i(\omega_B t + k_B \hat{y} + \phi_B)} \right] \right] dp_x dp_y, \quad (B5)$$

where we let $g_A = \langle 1|\bar{\varepsilon} \cdot \bar{E}_{A0}|2\rangle = \langle 2|\bar{\varepsilon} \cdot \bar{E}_{A0}|1\rangle$ and $g_B = \langle 3|\bar{\varepsilon} \cdot \bar{E}_{B0}|2\rangle = \langle 2|\bar{\varepsilon} \cdot \bar{E}_{B0}|3\rangle$. Finally, we can use the identities [39]:

$$e^{ik\hat{y}} = \sum_n \iint |n,p_x,p_y\rangle\langle n,p_x,p_y - \hbar k| dp_x dp_y \quad (B6a)$$

and

$$e^{-ik\hat{y}} = \sum_n \iint |n,p_x,p_y\rangle\langle n,p_x,p_y + \hbar k| dp_x dp_y, \quad (B6b)$$

and the rotating wave approximation [45] in Eq. (B5) to give:

$$H_1 = \iint \left[ \frac{\hbar g_A}{2} e^{i(\omega_A t + \phi_A)} |1,p_x,p_y\rangle\langle 2,p_x,p_y + \hbar k_A| \right.$$
$$\left. + \frac{\hbar g_A}{2} e^{-i(\omega_A t + \phi_A)} |2,p_x,p_y + \hbar k_A\rangle\langle 1,p_x,p_y| \right.$$

$$+\frac{\hbar g_B}{2}e^{i(\omega_B t+\phi_B)}|3,p_x,p_y+\hbar k_A+\hbar k_B\rangle\langle 2,p_x,p_y+\hbar k_A|$$

$$+\frac{\hbar g_B}{2}e^{-i(\omega_B t+\phi_B)}|2,p_x,p_y+\hbar k_A\rangle\langle 3,p_x,p_y+\hbar k_B+\hbar k_A|\bigg]dp_x\,dp_y. \quad (B7)$$

We note that the full interaction between the internal states $|1\rangle, |2\rangle,$ and $|3\rangle$ occurs across groups of three different momentum components: $|p_x,p_y\rangle, |p_x,p_y+\hbar k_A\rangle,$ and $|p_x,p_y+\hbar k_A+\hbar k_B\rangle$. This can be understood physically in terms of photon absorption/emission and conservation of momentum. Keeping in mind assumption (4), if an atom begins in state $|1,p_{x0},p_{y0}\rangle$ and absorbs a photon from field $\vec{E}_A$, it will transition to internal state $|2\rangle$ because it has become excited, but it will also gain the momentum of the photon ($\hbar k_A$) traveling in the $+y$ direction. It will therefore end up in state $|2,p_{x0},p_{y0}+\hbar k_A\rangle$. Now the atom is able to interact with field $\vec{E}_B$, which can cause stimulated emission of a photon with momentum $\hbar k_B$ in the $-y$ direction. If such a photon is emitted, the atom itself will gain an equal momentum in the opposite direction, bringing it into external state $|p_{x0},p_{y0}+\hbar k_A+\hbar k_B\rangle$. The atom will also make an internal transition to state $|3\rangle$ because of the de-excitation. The total state will now be $|3,p_{x0},p_{y0}+\hbar k_A+\hbar k_B\rangle$. We thereby see that our mathematics is corroborated by physical intuition.

Getting back to the Hamiltonian, we look at the general case of one momentum grouping so that we get in matrix form $H = H_0 + H_1$ from Eqs. (B3) and (B7):

$$H = \begin{bmatrix} \frac{p_x^2+p_y^2}{2m}+\hbar\omega_1 & \frac{\hbar g_A}{2}e^{i(\omega_A t+\phi_A)} & 0 \\ \frac{\hbar g_A}{2}e^{-i(\omega_A t+\phi_A)} & \frac{p_x^2+(p_y+\hbar k_A)^2}{2m}+\hbar\omega_2 & \frac{\hbar g_B}{2}e^{-i(\omega_B t+\phi_B)} \\ 0 & \frac{\hbar g_B}{2}e^{i(\omega_B t+\phi_B)} & \frac{p_x^2+(p_y+\hbar k_A+\hbar k_B)^2}{2m}+\hbar\omega_3 \end{bmatrix}. \quad (B8)$$

In order to remove the time dependence we apply some transformation $Q$ [39] of the form:

$$Q = \begin{bmatrix} e^{i(\theta_1 t+\phi_1)} & 0 & 0 \\ 0 & e^{i(\theta_2 t+\phi_2)} & 0 \\ 0 & 0 & e^{i(\theta_3 t+\phi_3)} \end{bmatrix}, \quad (B9)$$

so that the SE becomes

$$i\hbar \frac{\partial |\tilde{\Psi}\rangle}{\partial t} = \tilde{H}|\tilde{\Psi}\rangle, \tag{B10}$$

where $|\tilde{\Psi}\rangle = Q|\Psi\rangle$ and $\tilde{H} = QHQ^{-1} + i\hbar \frac{\partial Q}{\partial t} Q^{-1}$. The matrix representation is:

$$\tilde{H} = \begin{bmatrix} \frac{p_x^2 + p_y^2}{2m} + \hbar\omega_1 - \hbar\theta_1 & \frac{\hbar g_A}{2} e^{i(\omega_A + \theta_1 - \theta_2)t + i(\phi_A + \phi_1 - \phi_2)} & 0 \\ \frac{\hbar g_A}{2} e^{-i(\omega_A + \theta_1 - \theta_2)t - i(\phi_A + \phi_1 - \phi_2)} & \frac{p_x^2 + (p_y + \hbar k_A)^2}{2m} + \hbar\omega_2 - \hbar\theta_2 & \frac{\hbar g_B}{2} e^{-i(\omega_B + \theta_3 - \theta_2)t - i(\phi_B + \phi_3 - \phi_2)} \\ 0 & \frac{\hbar g_B}{2} e^{i(\omega_B + \theta_3 - \theta_2)t + i(\phi_B + \phi_3 - \phi_2)} & \frac{p_x^2 + (p_y + \hbar k_A + \hbar k_B)^2}{2m} + \hbar\omega_3 - \hbar\theta_3 \end{bmatrix}. \tag{B11}$$

Choosing $\theta_1 = -\omega_A, \theta_2 = 0, \theta_3 = -\omega_B$, $\phi_1 = -\phi_A$, $\phi_2 = 0$, and $\phi_3 = -\phi_B$, Eq. (B11) becomes:

$$\tilde{H} = \begin{bmatrix} E_1(p_x, p_y) + \hbar\omega_1 + \hbar\omega_A & \frac{\hbar g_A}{2} & 0 \\ \frac{\hbar g_A}{2} & E_2(p_x, p_y) + \hbar\omega_2 & \frac{\hbar g_B}{2} \\ 0 & \frac{\hbar g_B}{2} & E_3(p_x, p_y) + \hbar\omega_3 + \hbar\omega_B \end{bmatrix}, \tag{B12}$$

where we have taken:

$$\begin{aligned} E_1(p_x, p_y) &= \frac{p_x^2 + p_y^2}{2m}, \\ E_2(p_x, p_y) &= \frac{p_x^2 + (p_y + \hbar k_A)^2}{2m}, \\ E_3(p_x, p_y) &= \frac{p_x^2 + (p_y + \hbar k_A + \hbar k_B)^2}{2m}. \end{aligned} \tag{B13}$$

In order to further simplify the analysis, we set the zero energy at $E_1(p_{x0}, p_{y0}) + \hbar\omega_1 + \hbar\omega_A$ for some specific momentum group with $p_x = p_{x0}$ and $p_y = p_{y0}$. Also, since $\omega_A$ and $\omega_B$ can be chosen independently, we can let $E_3(p_{x0}, p_{y0}) + \hbar\omega_3 + \hbar\omega_B = E_1(p_{x0}, p_{y0}) + \hbar\omega_1 + \hbar\omega_A$. With the energies thus set, Eq. (B12) becomes:

$$\tilde{H} = \begin{bmatrix} 0 & \frac{\hbar g_A}{2} & 0 \\ \frac{\hbar g_A}{2} & -\delta & \frac{\hbar g_B}{2} \\ 0 & \frac{\hbar g_B}{2} & 0 \end{bmatrix}, \quad (B14)$$

where $\delta = (E_1(p_{x0}, p_{y0}) + \hbar\omega_1 + \hbar\omega_A) - (E_2(p_{x0}, p_{y0}) + \hbar\omega_2)$. Using this Hamiltonian in Eq. (B10), we get the equations of motion as:

$$\dot{\tilde{C}}_1(p_{x0}, p_{y0}, t) = -i\frac{g_A}{2}\tilde{C}_2(p_{x0}, p_{y0} + \hbar k_A, t),$$

$$\dot{\tilde{C}}_2(p_{x0}, p_{y0} + \hbar k_A, t) = -i\frac{g_A}{2}\tilde{C}_1(p_{x0}, p_{y0}, t) + i\delta\tilde{C}_2(p_{x0}, p_{y0} + \hbar k_A, t)$$
$$- i\frac{g_B}{2}\tilde{C}_3(p_{x0}, p_{y0} + \hbar k_A + \hbar k_B, t), \quad (B15)$$

$$\dot{\tilde{C}}_3(p_{x0}, p_{y0} + \hbar k_A + \hbar k_B, t) = -i\frac{g_B}{2}\tilde{C}_2(p_{x0}, p_y + \hbar k_A, t).$$

Assumption (5) allows us to make the adiabatic approximation so that we can set $\dot{\tilde{C}}_2(p_{x0}, p_{y0}, t) \approx 0$, and assumption (6) gives us $g_A = g_B = g_0$. The Eqs. (B15) then simplify to:

$$\dot{\tilde{C}}_1(p_{x0}, p_{y0}, t) = -i\frac{g_0^2}{4\delta}\tilde{C}_1(p_{x0}, p_{y0}, t) - i\frac{g_0^2}{4\delta}\tilde{C}_3(p_{x0}, p_{y0} + \hbar k_A + \hbar k_B, t),$$

$$\dot{\tilde{C}}_3(p_{x0}, p_{y0} + \hbar k_A + \hbar k_B, t) = -i\frac{g_0^2}{4\delta}\tilde{C}_1(p_{x0}, p_{y0}, t) - i\frac{g_0^2}{4\delta}\tilde{C}_3(p_{x0}, p_{y0} + \hbar k_A + \hbar k_B, t), \quad (B16)$$

where we have chosen to neglect state $C_2$ from here on due to the adiabatic approximation. We can now use another transformation on this system to make it more tractable. Let:

$$\tilde{\tilde{C}}_1(p_{x0}, p_{y0}, t) = \tilde{C}_1(p_{x0}, p_{y0}, t) e^{i\frac{g_0^2}{4\delta}t},$$

$$\tilde{\tilde{C}}_3(p_{x0}, p_{y0} + \hbar k_A + \hbar k_B, t) = \tilde{C}_3(p_{x0}, p_{y0} + \hbar k_A + \hbar k_B, t) e^{i\frac{g_0^2}{4\delta}t}.$$

(B17)

.

The system in Eqs. (B16) then becomes:

$$\dot{\tilde{\tilde{C}}}_1(p_{x0}, p_{y0}, t) = -i\frac{g_0^2}{4\delta} \tilde{\tilde{C}}_3(p_{x0}, p_{y0} + \hbar k_A + \hbar k_B, t),$$

$$\dot{\tilde{\tilde{C}}}_3(p_{x0}, p_{y0} + \hbar k_A + \hbar k_B, t) = -i\frac{g_0^2}{4\delta} \tilde{\tilde{C}}_1(p_{x0}, p_{y0}, t).$$

(B18)

Solving this and reversing the transformations of Eqs. (B17) and (B9), we arrive at:

$$C_1(p_{x0}, p_{y0}, t) = C_1(p_{x0}, p_{y0}, 0) \cos\left(\frac{\Omega}{2}t\right)$$
$$- i e^{i\left(\omega_A - \frac{\Omega}{2}\right)t + i\phi_A} \left[ C_3(p_{x0}, p_{y0} + \hbar k_A + \hbar k_B, 0) e^{-i\left(\omega_B - \frac{\Omega}{2}\right)t - i\phi_B} \sin\left(\frac{\Omega}{2}t\right) \right],$$

(B19)

$$C_3(p_{x0}, p_{y0} + \hbar k_A + \hbar k_B, t) = -i e^{i\left(\omega_B - \frac{\Omega}{2}\right)t + i\phi_B} \left[ C_1(p_{x0}, p_{y0}, 0) e^{-i\left(\omega_A - \frac{\Omega}{2}\right)t - i\phi_A} \sin\left(\frac{\Omega}{2}t\right) \right]$$
$$+ C_3(p_{x0}, p_{y0} + \hbar k_A + \hbar k_B, 0) \cos\left(\frac{\Omega}{2}t\right),$$

where we let $\Omega = \frac{g_0^2}{2\delta}$. It should be noted, however, that these solutions were arrived at only for the specific momentum group where $p_x = p_{x0}$ and $p_y = p_{y0}$. This was the case where both laser fields were equally far detuned. Other momentum groups will have slightly different solutions due to the Doppler shift, which causes the detunings to be perturbed. For a more accurate description, we need to numerically solve each momentum group's original three equations of motion without making any approximations. This is what we do in our computational model. For a basic phenomenological understanding of the interferometer, however, it is sufficient to assume that the above analytical solution is accurate for all momentum components

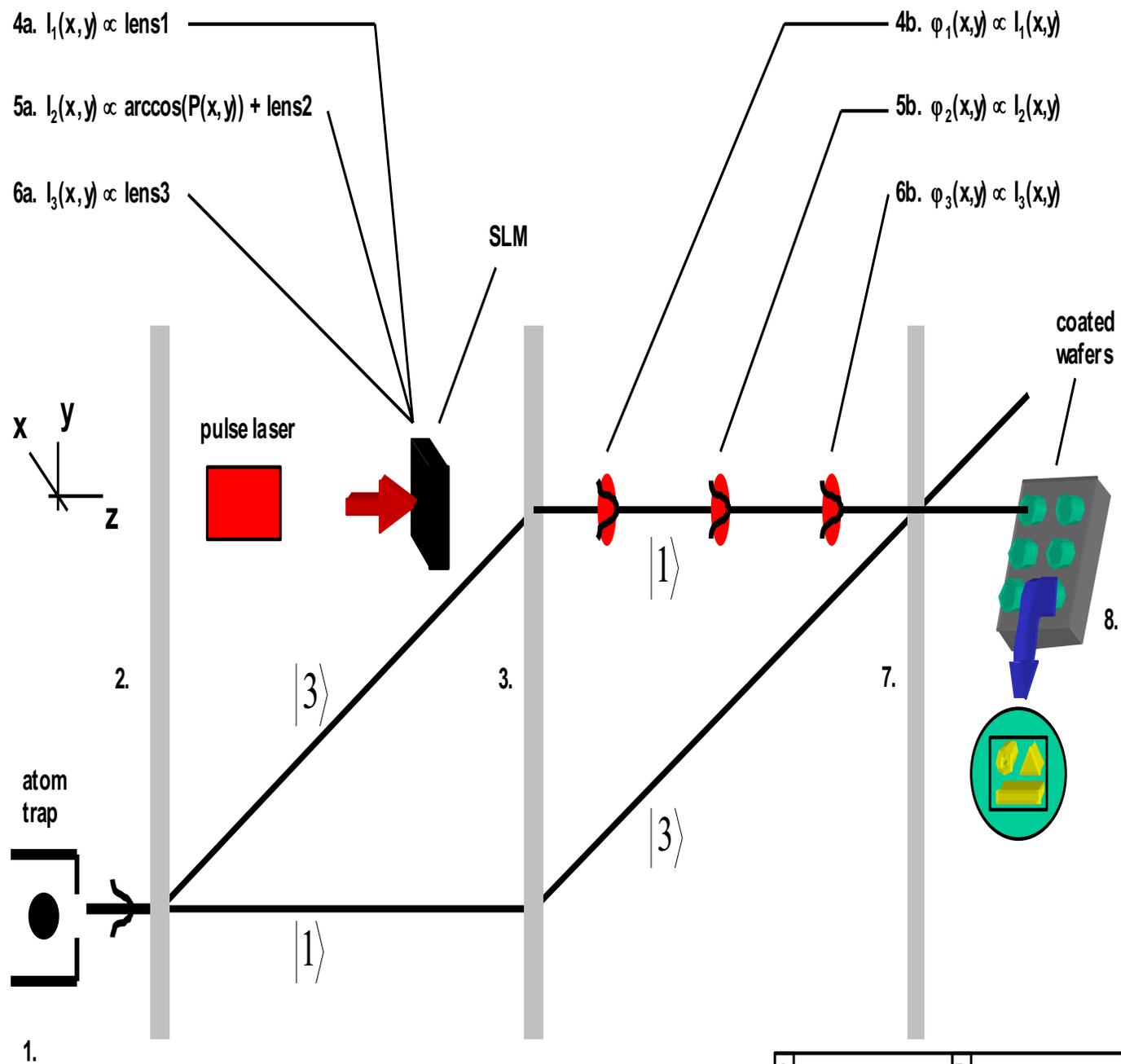
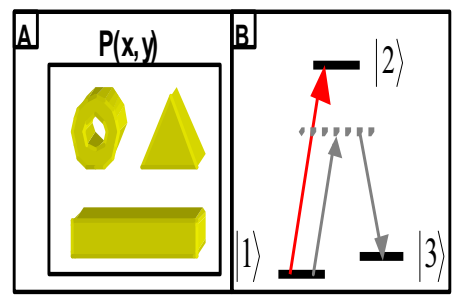

4a. $I_1(x,y) \propto$ lens1

5a. $I_2(x,y) \propto \arccos(P(x,y)) +$ lens2

6a. $I_3(x,y) \propto$ lens3

4b. $\varphi_1(x,y) \propto I_1(x,y)$

5b. $\varphi_2(x,y) \propto I_2(x,y)$

6b. $\varphi_3(x,y) \propto I_3(x,y)$

SLM

pulse laser

coated wafers

atom trap

$|1\rangle$ $|3\rangle$ $|3\rangle$ $|1\rangle$

1. 2. 3. 7. 8.

A  P(x,y)   B  $|2\rangle$  $|1\rangle$  $|3\rangle$

FIG. 1. 1.) A single atomic wavepacket is released from the atom trap. 2.) The wavepacket is split using a π/2 pulse. 3.) The split components are reflected by a π pulse. 4a.) The spatial light modulator (SLM) modulates a light pulse such that it will act as the first lens of the atom lens system. 4b.) The light pulse intercepts the wavepacket component that is in state $|1\rangle$ and imparts a phase signature $\phi_1(x,y)$ via the ac-stark effect. 5a.) Now the SLM modulates a second light pulse such that it will impart the both the phase information corresponding to the arbitrary image (arcos(P(x,y))) and the phase information of the second lens of the lens system. 5b.) The second light pulse intercepts the same wavepacket component as the first one and imparts the phase signature $\phi_2(x,y)$. 6a.) The SLM modulates a third light pulse, preparing it to act as the third lens of the lens system. 6b.) The third light pulse intercepts the same wavepacket component as the other two pulses and imparts a phase $\phi_3(x,y)$. 7.) Both wavepacket components are mixed along the two trajectories by a π/2 pulse. 8.) A chemically treated wafer intercepts the state $|1\rangle$ component so that an interference pattern forms on the wafer proportional to 1+cos[arcos(P(x,y))] = 1 + P(x,y). INSET A. The image P(x,y) that is to be transferred ultimately to the wafer. INSET B. The internal energy states of the wavepacket modeled as a lambda system. The light pulses used for the atom lenses have a much larger detuning for ground state $|3\rangle$ than they do for ground state $|1\rangle$ so that they effectively only interact with the state $|1\rangle$ component of the wavepacket. The π/2 pulses and the π pulse use light that is largely detuned for both ground states.

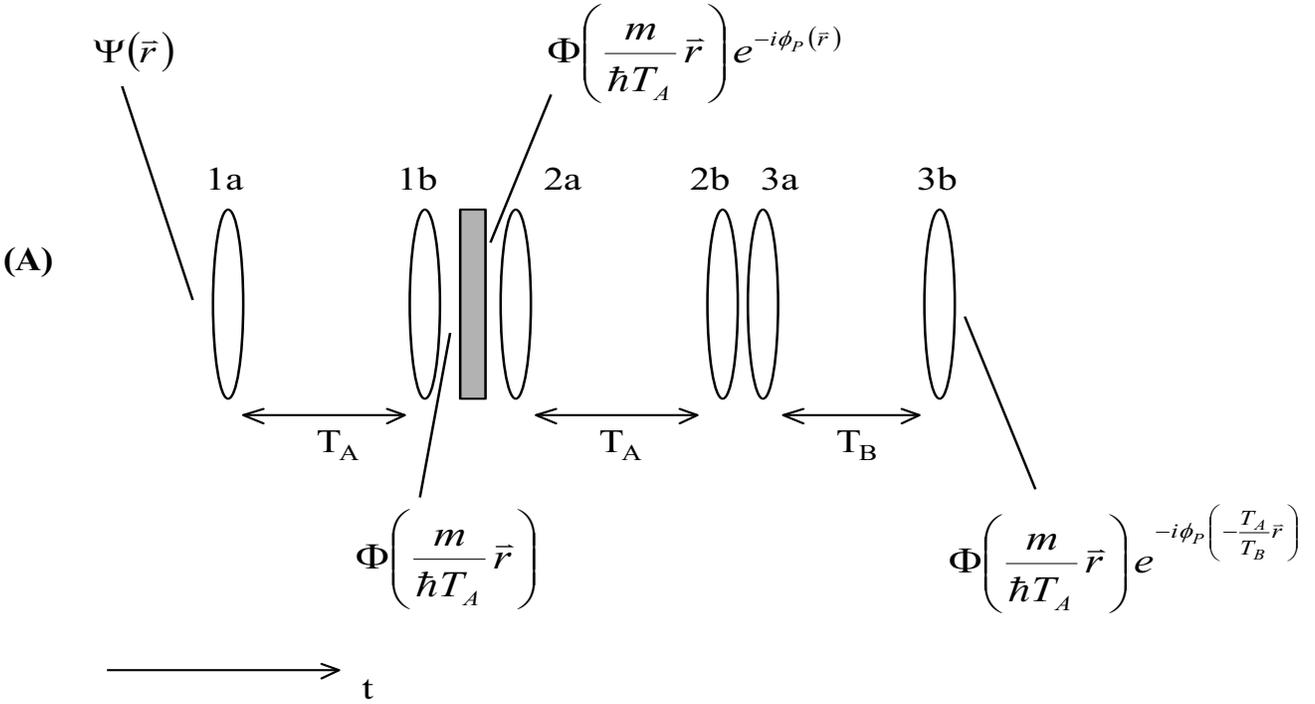

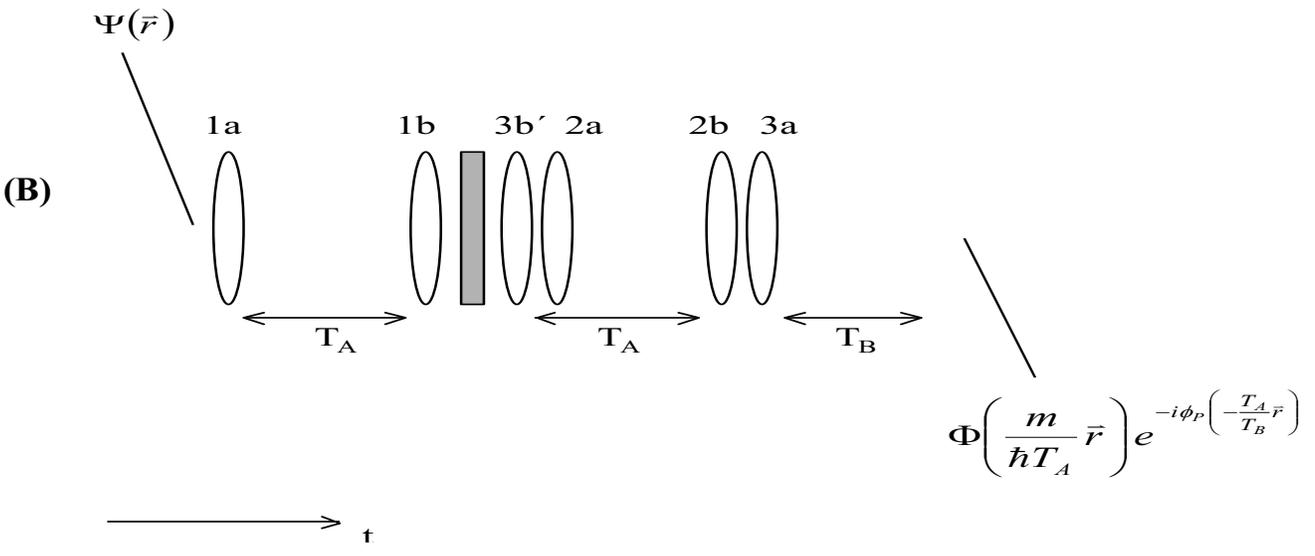

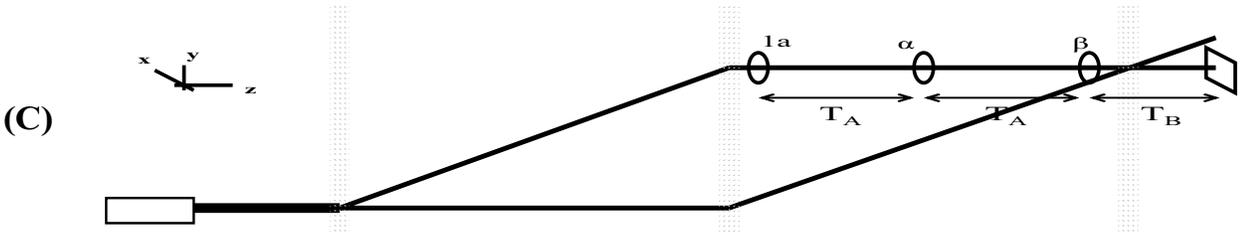

FIG. 2. (A) The lens system. Each lens is actually a pulse of light with a transverse intensity modulation. Between lenses 1a and 1b and 2a and 2b are freespace regions of time duration $T_A$, while between lenses 3a and 3b there is a freespace region of duration $T_B$. Lenses 1a and 2a give the wavefunction a phase $\phi_{1a} = \phi_{2a} = -\frac{m}{2\hbar T_A}|\vec{r}|^2$, lenses 1b and 2b impart a phase $\phi_{1b} = \phi_{2b} = -\frac{m}{2\hbar T_A}|\vec{r}|^2 + \frac{\pi}{2}$, lens 3a gives a phase $\phi_{3a} = -\frac{m}{2\hbar T_B}|\vec{r}|^2$, and lens 3b gives a phase $\phi_{3b} = -\frac{m}{2\hbar T_B}|\vec{r}|^2 + \frac{\pi}{2}$. (B) The lens system from (A) rearranged. The input and output are still the same, but the output is no longer immediately preceded by a lens. Lens 3b' is the same as lens 3b from (A) except for a $T_A$ in place of $T_B$ so that it gives a phase shift of $\phi_{3b'} = -\frac{m}{2\hbar T_A}|\vec{r}|^2 + \frac{\pi}{2}$. (C) The modified lens system in context. The π and π/2 pulses are not shown for the sake of simplicity. Lenses α and β are composites of the lenses from the system of (B). Between lenses 1a and α is a freespace region of time length $T_A$, as well as between lenses α and β. Between lens β and the substrate is a freespace region of time duration $T_B$. $\phi_\alpha(\vec{r}) = -\left(\frac{3m}{2\hbar T_A}\right)|\vec{r}|^2 + \pi - \phi_P(\vec{r})$,

$\phi_\beta(\vec{r}) = -\left(\frac{m}{2\hbar T_A} + \frac{m}{2\hbar T_B}\right)|\vec{r}|^2 + \frac{\pi}{2}$.

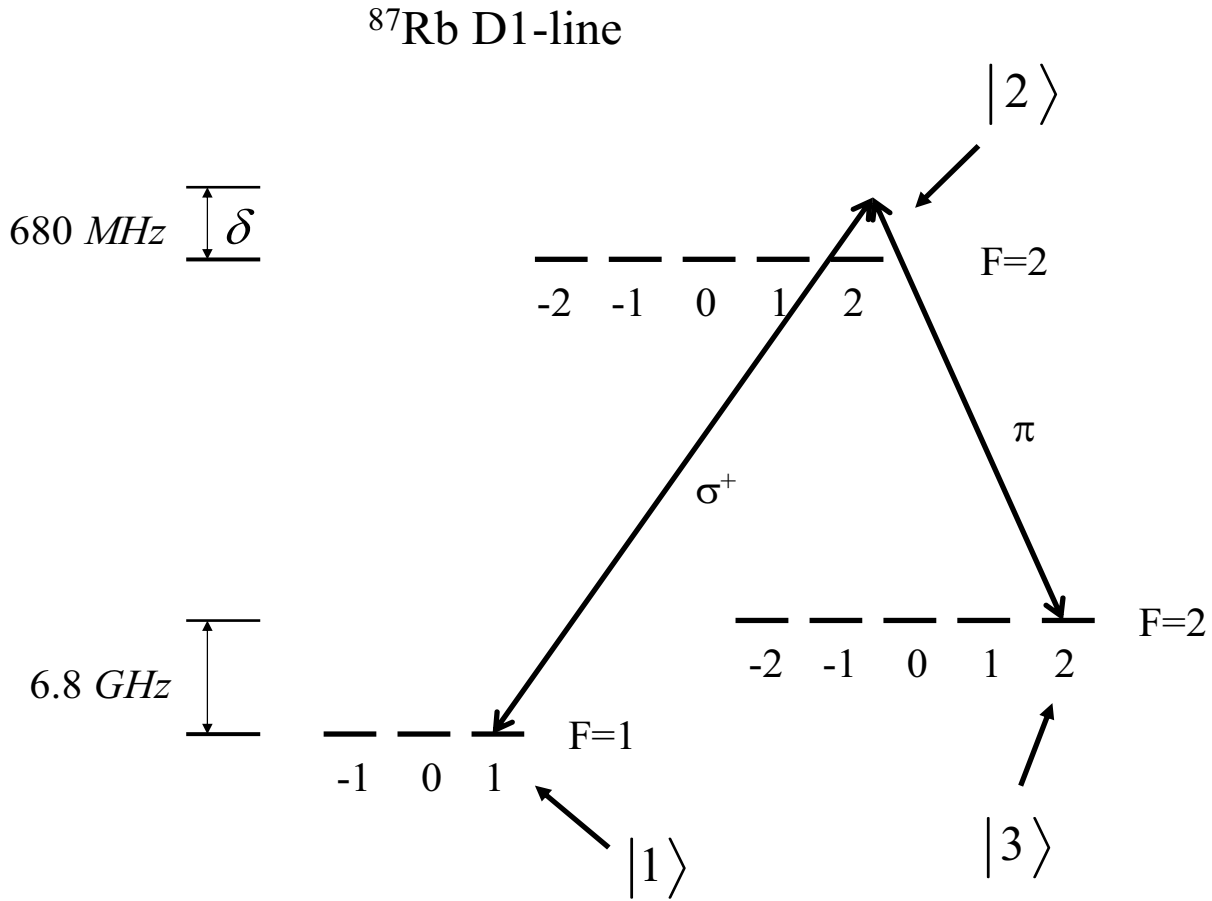

**FIG. 3.** The transition scheme. $\sigma^+$-polarized light excites the $|1\rangle \leftrightarrow |2\rangle$ transition and $\pi$-polarized light excites the $|2\rangle \leftrightarrow |3\rangle$ transition. Both lasers are detuned by 680 MHz. For the $\pi$ and $\pi/2$ pulses, the two transitions are simultaneously excited. For the light shift based lens system, only the $\pi$-polarized light is applied so as to affect only state $|3\rangle$. The detuning of the lasers is small enough such that all other transitions from the states $|1\rangle$, $|2\rangle$, and $|3\rangle$ that are excitable by either the $\sigma^+$-polarized or $\pi$-polarized light see a detuning that is at least a factor of 10 larger. They can therefore be neglected.

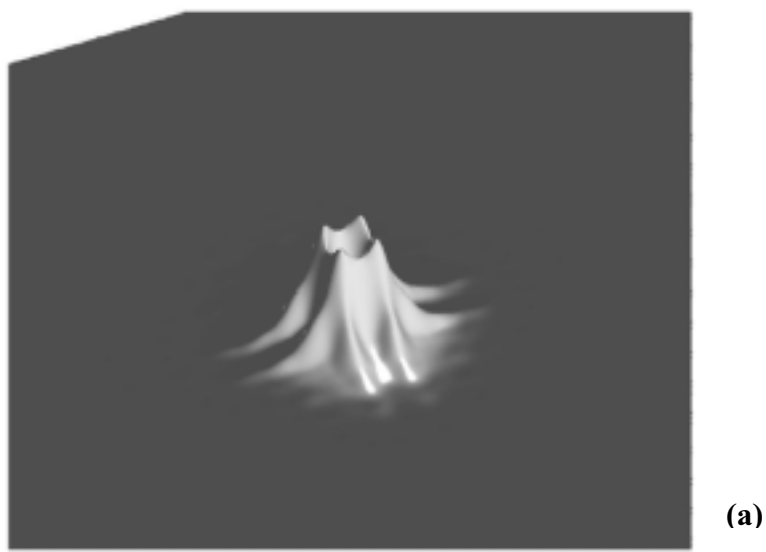

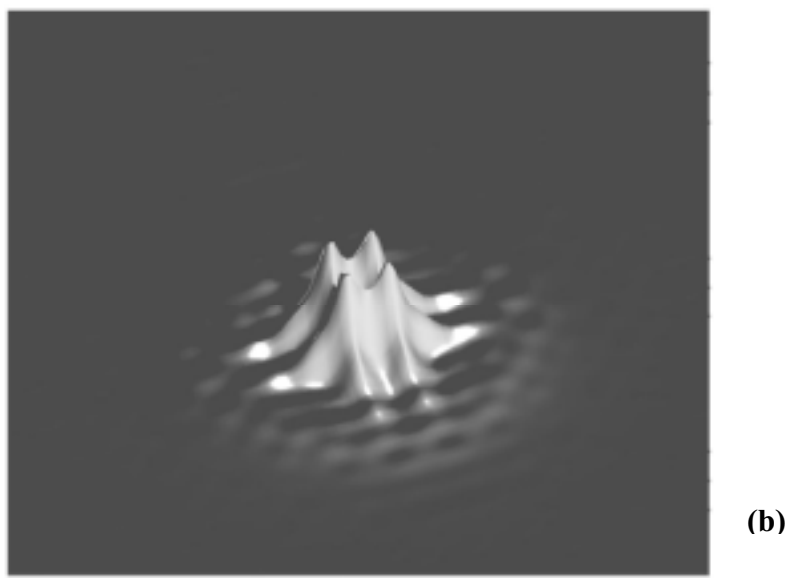

**FIG. 4.** (a) An arbitrary image is formed with the lens system in place, but without any scaling. We see that it is a more complex pattern than just a simple periodic structure such as sinusoidal fringes.(b) The same image as in (a) is formed with the lens system still in place, but a scaling factor of two has been used to shrink the pattern.